\documentstyle[sprocl]{article}

\def\NPB{{\em Nucl. Phys.} B}
\def\PLB{{\em Phys. Lett.}  B}
\def\PRL{\em Phys. Rev. Lett.}

\def\be{\begin{equation}}
\def\ee{\end{equation}}
\def\bea{\begin{eqnarray}}
\def\eea{\end{eqnarray}}
\def\NPB#1#2#3{Nucl. Phys. B{#1} (19#2) #3}
\def\PLB#1#2#3{Phys. Lett. B{#1} (19#2) #3}

\def\PRL#1#2#3{Phys. Rev. Lett. {#1} (19#2) #3}

\def\yzero{\smash{\hbox{$y\kern-4pt\raise1pt\hbox{${}^\circ$}$}}}

\def\beq{\begin{equation}}  
\def\eeq{\end{equation}}
\def\beqa{\begin{eqnarray}}
\def\eeqa{\end{eqnarray}}

\def\-{\hphantom{-}}

\def\s2{\frac{1}{\sqrt2}}

\def\beq{\begin{equation}}
\def\eeq{\end{equation}}
\def\beqa{\begin{eqnarray}}
\def\eeqa{\end{eqnarray}}

\def\IF{\relax{\rm I\kern-.18em F}}
\def\II{\relax{\rm I\kern-.18em I}}
\def\IP{\relax{\rm I\kern-.18em P}}
\def\IC{\relax\hbox{\kern.25em$\inbar\kern-.3em{\rm C}$}}
\def\IR{\relax{\rm I\kern-.18em R}}

\def\Dsl{\,\raise.15ex\hbox{/}\mkern-13.5mu D} %this one can be subscripted
\def\IZ{Z\kern-.4em  Z}

\catcode`\@=11
\newdimen\@rotdimen 
\newbox\@rotbox

\def\@vspec#1{\special{ps:#1}}%  passes #1 verbatim to the output
\def\@rotstart#1{\@vspec{gsave currentpoint currentpoint translate
   #1 neg exch neg exch translate}}% #1 can be any origin-fixing
\def\@rotfinish{\@vspec{currentpoint grestore moveto}}% gets back in synch
\def\@rotr#1{\@rotdimen=\ht#1\advance\@rotdimen by\dp#1%
   \hbox to\@rotdimen{\hskip\ht#1\vbox to\wd#1{\@rotstart{90 rotate}%
   \box#1\vss}\hss}\@rotfinish}
\def\@rotl#1{\@rotdimen=\ht#1\advance\@rotdimen by\dp#1%
   \hbox to\@rotdimen{\vbox to\wd#1{\vskip\wd#1\@rotstart{270 rotate}%
   \box#1\vss}\hss}\@rotfinish}%
\def\@rotu#1{\@rotdimen=\ht#1\advance\@rotdimen by\dp#1%
   \hbox to\wd#1{\hskip\wd#1\vbox to\@rotdimen{\vskip\@rotdimen
   \@rotstart{-1 dup scale}\box#1\vss}\hss}\@rotfinish}%
\def\@rotf#1{\hbox to\wd#1{\hskip\wd#1\@rotstart{-1 1 scale}%
   \box#1\hss}\@rotfinish}%
\def\rotate{\@ifnextchar[{\@rotate}{\@rotate[l]}}
\def\@rotate[#1]#2{\setbox\@rotbox=\hbox{#2}\@nameuse{@rot#1}\@rotbox}

\catcode`\@=12

\begin{document}
\begin{flushright}
{IFT-UAM/CSIC-01-25 ; FTUAM-01/17}\\
%%\date{\today}
\end{flushright}

\vspace{1cm}
\title{Standard Model Engineering with Intersecting Branes
\footnote{To appear in the proceedings of SUSY-01, Dubna (Russia),  June
2001.}}

\author{Luis E. Ib\'a\~nez}

\address{Departamento de F\'{\i}sica Te\'orica C-XI
and Instituto de F\'{\i}sica Te\'orica  C-XVI,\\
Universidad Aut\'onoma de Madrid,
Cantoblanco, 28049 Madrid, Spain }

\maketitle\abstracts{ 
I briefly describe the recent construction of 
the first specific string D-brane models 
yielding just the SM massless fermion spectrum. One of the
most remarkable facts predicted
by these constructions is that the global symmetries of the 
SM like Baryon and Lepton numbers as well as the axial $U(1)_A$
 are gauged symmetries whose  anomalies are cancelled by a Green-Schwarz
mechanism only in the case of  three quark-lepton generations.  
Proton stability and the Dirac nature of 
neutrino masses  are thus a prediction of these theories. 
This is work done in collaboration with Fernando Marchesano and
Raul Rabad\'an.}

%\section{Guidelines}
%\subsection{Producing the Hard Copy}\label{subsec:prod}

%\section{Introduction}

Since 1984 a lot of effort has been devoted to the construction of
string compactifications yielding a spectrum as close as possible
to the Standard Model (SM). A number of explicit three-generation
models have been constructed and other important phenomenological 
questions have been addressed. In spite of these successes  
 I think it is fair to say that 
up to now there is not a completely satisfactory model. The models
in general have a tendency to produce extra unwanted gauge interactions 
and extra fermions beyond the minimal content of the SM (or the MSSM). 
Although some models are argued to get closer to the observed physics
their analysis requires a very specific and arbitrary choice of 
scalar flat directions. Most importantly, crucial properties
like proton stability depend on these  particular choices of
scalar vacua.

One of the important difficulties in building realistic models
from the perturbative heterotic string was the necessity of following
a {\it top-bottom} approach to model building: one starts from
a theory in higher dimensions and a large gauge group 
(e.g., $E_8\times E_8$) and breaks the symmetry down to 
$SU(3)\times SU(2)\times U(1)$ as well as one can. This procedure
has as a consequence the proliferation of extra fermions and 
gauge symmetries beyond the searched SM.

Our new understanding of string theory has changed the situation drastically.
We  know now that the Dp-branes,  solitonic sectors  of Type II string theory,
are the natural objects when discussing gauge interactions. Dp-branes
have  gauge (and matter) particles living on a $(p+1)$ submanifold of the whole 
10 (or 11) dimensions. A set of $N$ parallel Dp-branes generically have a
$U(N)$ 
gauge theory living on their worldvolume.
 Thus we have now the ingredients to follow a
{\it bottom-up approach} to the construction of a realistic string model:
we just have to consider some stacks of D-branes for the different
SM gauge groups $SU(3)$, $SU(2)$ and $U(1)$.
Unfortunately life is not so easy: the  gauge theories on the worldvolume
of Dp-branes are {\it non-chiral}. For example, $M$ parallel D3-branes
yield a 4-dimensional gauge theory with  $N=4$ supersymmetry.

In order to get a chiral theory on the worldvolume of D-branes there are
not that many options known at the moment. One of the simplest is 
locating the (9-p) transverse coordinates of the considered  brane
on top of some singularity in extra dimensions
\cite{singus}. For example, locating
D3-branes on top of an orbifold $C^3/Z_N$ singularity. This system
of branes can be embedded in a variety of manners in a compact 
6-dimensional space \cite{aiqu,evenmore} and one can get semi-realistic models 
with three generations of quarks and leptons of the SM or 
 left-right symmetric models. This new classes of models are 
interesting in their own right and provide explicit examples 
of semi-realistic brane-world theories with gravity mediated SUSY-breaking  
or 1 TeV brane world models with SM group and three generations
\cite{evenmore,aiqu,angel,bail,bjl}.

The other known alternative in order to get chiral fermions on the 
worldvolume of D-branes is intersecting branes
\cite{bdl}. Consider for example 
a bunch of N Dp-branes and another set of M Dp-branes ($p>3$) containing
both
Minkowski space and intersecting at some angle  in the
(p-3) extra dimensions. Then one can check that in general one
will get massless chiral fermions transforming like 
$(N,{\overline M})$ under the gauge group $U(N)\times U(M)$. 
This is quite interesting because the SM is a product of unitary groups 
and quarks and leptons may be considered  to transform under 
bifundamental representations. In addition there is a very 
interesting property: if the extra 6 dimensions are compact, the intersection
of a couple of branes is in general multiple
\cite{bgkl,afiru,afiru2}. Thus in the 
context of intersecting branes the replication of generations is natural.
This structure to get a SM spectrum is quite general. Some simple examples
may be obtained by considering D(3+n) branes wrapping on the compact space
$T^{2n}\times (T^{2(3-n)}/Z_N)$ , for $n=1,2,3$ and intersecting at angles 
in the $T^{2n}$ (see
refs.\cite{bgkl,afiru,afiru2,bkl,imr,bklo,csu,bailin}). 

Recently a particularly interesting class of models yielding just
the massless fermion spectrum of the SM was constructed
\cite{imr}. Let me
summarize the main features of these constructions. We consider 
Type IIA  string theory compactified on a six-torus  
$T^2\times T^2\times T^2$.
Because of technical reasons to be explained below we do
an "orientifold" twist 
\cite{orientold} of this theory by the product
$\Omega \times R$, where $\Omega$ is the worldsheet parity operator
and $R$ is the reflection operator with respect of one of the axis 
of each tori. Now we consider D6-branes containing inside Minkowski
space and wrapping each of the three remaining dimensions 
\cite{bgkl}of the
branes on a different torus $T^2$. We will denote by $(n_a^i,m_a^i)$, i=1,2,3
the wrapping numbers of each brane $D6_a$, $n_a^i$($m_a^i$) being the number 
of times the brane is wrapping around the  x(y)-coordinate of the $i-th$
torus. One can check that the number of times two branes 
$D6_a$  and $D6_b$ intersect in $T^6$ is given by the intersection number
\cite{bgkl}:
\beq
I_{ab}\ =\
(n_a^1m_b^1-m_a^1n_b^1)(n_a^2m_b^2-m_a^2n_b^2)(n_a^3m_b^3-m_a^3n_b^3)
\label{internumber}
\eeq
Open strings stretching around the intersections give rise to 
chiral fermions in the bifundamental representation $(N_a,{\overline N}_b)$
under the gauge group of the two branes $U(N_a)\times U(N_b)$. 
Thus these configurations  yield $I_{ab}$ copies of the same
bifundamental representation, providing a natural source for the observed
generation replication. Let us now show how to build a configuration
giving just three generations of a $SU(3)\times SU(2)\times U(1)_Y$ 
group \cite{imr}. We will consider four stacks of branes with multiplicities
$N_a=3$, $N_b=2$, $N_c=1$, $N_d=1$, giving thus rise to an
initial gauge group $U(3)\times U(2)\times U(1)\times U(1)$
\footnote{Although apparently such a structure would yield four gauged
$U(1)$'s, we show below that we expect three of these $U(1)$'s to
become massive and decouple from the low-energy spectrum.}.
Now, it is easy to find wrapping numbers 
$(n_a^i,m_a^i)$ for the four sets of D6-branes yielding the 
desired SM spectrum. It is enough to select them in such a way that 
the intersection numbers $I_{ij}$, $i,j=a,b,c,d$ are given by \cite{imr} 
\beqa
I_{ab}\ & = &  \ 1 \ ;\ I_{ab*}\ =\ 2  \nonumber \\
I_{ac}\ & = &  \ -3 \ ;\ I_{ac*}\ =\ -3  \nonumber \\
I_{bd}\ & = &  \ 0  \ ;\ I_{bd*}\ =\ -3  \nonumber \\
I_{cd}\ & = &  \ -3 \ ;\ I_{cd*}\ =\ 3
\label{intersec2}
\eeqa
all other intersections vanishing.
Here a negative number denotes that the
corresponding fermions should have opposite chirality to those
with positive intersection number.   
The massless fermion spectrum living at the intersections  is
shown in  Table 1, as well as the charges with respect to the
four $U(1)$'s.
Note that, due to the orientifold operation $\Omega R$ one has to include
the D6-branes which are "mirror" under that operation and have the same
wrapping numbers except for a flip in sign for the $m_a^i$'s. 
We denote the
mirror D6-branes with a star.
It turns out that the intersection of a $D6_a$ brane with the 
mirror of a $D6_b$ brane gives rise to a bifundamental 
of the type $(N_a,N_b)$ ({\it not} $(N_a,{\overline N}_b)$). 
 Note that the left-handed quarks in the table 
are not universal under the $U(1)_b$ charge. This is because
one of them comes from the intersection $(ab)$ whereas the other
two come from intersections $(ab^*)$.
 This is forced upon us if we
want to have a consistent theory and this is why we did the
orientifold twist to begin with.
In all D-brane models strong constraints appear from
 cancellation of Ramond-Ramond (RR) tadpoles which in turn
 guarantees the
cancellation of gauge
anomalies. In the case of the D-brane models here discussed
tadpole  cancellation  requires that there should be
as many $N_a$ as ${\overline N}_a$ representations for any $U(N_a)$
group. In our particular case this lack of universality of 
left-handed quarks is required in order to have as many $U(2)$ doublets
as anti-doublets, as the reader may easily check.
 Note also that they cancel because
the number of generations  is equal to the number of colours \cite{imr}.
This connection between number of colours and numbers of generations in this
class of models is quite appealing.  
Note also that the above constraints also require the presence of 
right-handed neutrinos.

\begin{table}[htb] \footnotesize
\renewcommand{\arraystretch}{1.25}
\begin{center}
\begin{tabular}{|c|c|c|c|c|c|c|c|}
\hline Intersection &
 Matter fields  &   &  $Q_a$  & $Q_b $ & $Q_c $ & $Q_d$  & Y \\
\hline\hline (ab) & $Q_L$ &  $(3,2)$ & 1  & -1 & 0 & 0 & 1/6 \\
\hline (ab*) & $q_L$   &  $2( 3,2)$ &  1  & 1  & 0  & 0  & 1/6 \\
\hline (ac) & $U_R$   &  $3( {\bar 3},1)$ &  -1  & 0  & 1  & 0 & -2/3 \\
\hline (ac*) & $D_R$   &  $3( {\bar 3},1)$ &  -1  & 0  & -1  & 0 & 1/3 \\
\hline (bd*) & $ L$    &  $3(1,2)$ &  0   & -1   & 0  & -1 & -1/2  \\
\hline (cd) & $E_R$   &  $3(1,1)$ &  0  & 0  & -1  & 1  & 1   \\
\hline (cd*) & $N_R$   &  $3(1,1)$ &  0  & 0  & 1  & 1  & 0 \\
\hline \end{tabular}
\end{center} \caption{ Standard model spectrum and $U(1)$ charges
\label{tabpssm} }
\end{table}

In ref.\cite{imr}  we gave a general solution for the wrapping numbers
$(n_a^i,m_a^i)$
giving rise to a SM spectrum. In here we will limit ourselves to  a
specific example for illustrative purposes. Consider the  wrapping numbers given
in Table 2 for our four SM branes
\footnote{The effective fractional $m_a^3$ winding numbers in the
third torus come from the addition of a NS B-field background which is required
to get odd number of generations in the orientifold case \cite{bkl}.}
The reader can easily check
 using eq.(\ref{internumber}) that
these wrapping numbers give rise to the intersection numbers 
in eq.(\ref{intersec2}) and hence to the SM spectrum of Table 1.

\begin{table}[htb] \footnotesize
\renewcommand{\arraystretch}{2.5}
\begin{center}
\begin{tabular}{|c||c|c|c|}
\hline
 $N_i$    &  $(n_i^1,m_i^1)$  &  $(n_i^2,m_i^2)$   & $(n_i^3,m_i^3)$ \\
\hline\hline $N_a=3$ & $(1,0)$  &  $(2,1)$ &
 $(1 ,  1/2)$  \\
\hline $N_b=2$ &   $(0,-1)$    &  $ (1 ,0)$  &
$(1,3 /2)$   \\
\hline $N_c=1$ & $(1,3)$  &
 $(1,0)$  & $(0,1)$  \\
\hline $N_d=1$ &   $(1,0)$    &  $(0,-1 )$  &
$(1, 3/2)$   \\
\hline \end{tabular}
\end{center} \caption{Example of  D6-brane wrapping numbers giving rise to a SM
spectrum.
\label{solution} }
\end{table}

The analysis of the $U(1)$ gauge symmetries in these models is quite interesting.
The four $U(1)$ symmetries $Q_a$, $Q_b$, $Q_c$ and $Q_d$ have clear
interpretations in terms of known global symmetries of the standard model.
Indeed $Q_a$ is $3B$, $B$ being the baryon number and $Q_d$ is nothing
but (minus)lepton number. Concerning $Q_c$, it is twice $I_R$, the third
component of right-handed weak isospin familiar from left-right
symmetric models. Finally $Q_b$ has the properties of a Peccei-Quinn
symmetry (it has mixed $SU(3)$ anomalies).
We thus learn that {\it all these known global symmetries of the SM
are in fact gauge symmetries} in this class of theories.
This is important because it implies that {\it Baryon and Lepton 
symmetries are gauged symmetries} in this class of theories.
As a consequence protons  are perturbatively stable and neutrino masses
must be Dirac \footnote{Notice this implies that cosmological
baryogenesis    
can only happen at the non-perturbative level,
as in weak-scale baryogenesis scenarios.}.

Two out of the four $U(1)$'s have triangle 
anomalies which are cancelled by a generalized Green-Schwarz mechanism
\cite{dsw,iru}.
The anomalous symmetries are  the $Q_b$ generator
and the $3Q_a-Q_d$ 
generator  which can be identified with the $(3B+L)$.
In these toroidal models there are four two-index antisymmetric 
tensors $B_2^{k}$, $k=0,1,2,3$ which come from the Ramond-Ramond closed 
string sector of the theory. They have $B_2\wedge F^j$ type couplings
with the four $U(1)$'s of the theory. Those can be computed and in the example 
above they are as follows \cite{imr}:
\beqa
B_2^1 &\wedge & \ (-2)\ F^{b} \nonumber \\
B_2^2 &\wedge  & \ 
3\ (3F^a\ -\  F^{d})
\nonumber \\
 B_2^3 &\wedge  &  \ (3\  F^a\ +\ 
F^c ) 
\label{bfs}
\eeqa
whereas the $B_2^0$ RR field has no couplings to the $F^j$.
 The antisymmetric $B_2^k$ fields are dual in four dimensions 
to pseudoscalars $C_0^k$. The duals to $B_2^1$ and
$B_2^2$  have couplings to the
full gauge group of the form:
\beqa
{1\over 2} &  C_0^1\ & (F^a\wedge F^a\ -\ 3F^d\wedge
F^d)\nonumber \\
{1\over 2 } & C_0^2\ &   (-F^b\wedge
F^b\ +\
2F^c\wedge F^c) 
\label{cff}
\eeqa
whereas the RR scalar $C_0^3$ does not couple to any $F\wedge F$ term.
It is easy to check that the combination of these couplings cancel 
all $U(1)$ anomalies by the tree-level exchange of the $B_2^k(C_0^k)$
$k=1,2$ fields in the usual way.
At the same time, the first two $B\wedge F$ couplings in eq.(\ref{bfs})
give a mass of order the string scale $M_s$  to the gauge bosons corresponding
to the anomalous combinations : $Q_b$ and $(3Q_a-Q_d)$. Indeed,
after a duality transformation the $B\wedge F$ couplings turn into 
Higgs-like couplings of type $M_s^2(\partial_{\mu }C_o^k) A_a^{\mu }$ 
which render those $U(1)$'s massive. In addition something 
interesting also happens \cite{imr}: the third linear combination
$(3Q_a+Q_c)$
which is anomaly-free {\it also becomes massive} by swallowing the
$B_2^3$ field in a similar way
\footnote{We are used to the fact that anomalous $U(1)$'s become
massive in D=4 string models. This is something new: now we see that 
anomaly-free $U(1)$'s may also become  massive due to the
generic presence of $B\wedge F$ couplings. 
Note also that  in the case of a low string scale $M_s$  
the physical SM $Z^0$ may owe  part of its mass to
a $B\wedge F^{Z^0}$ term due to a mixing of order
$M_W/M_s$ with the other $U(1)$'s, both anomalous
and non-anomalous  \cite{iq}. }
. Thus at the end of the day only
the  orthogonal linear combination $Q_Y=(1/6)Q_a-(1/2)Q_c+(1/2)Q_d$, which
is nothing but hypercharge,  remains massless
\footnote{The massive $U(1)$'s remain however as effective global
symmetries.}. Thus, as
promised, the
actual gauge group of the theory is 
just  $SU(3)\times SU(2)\times U(1)_Y$.

These D6-brane models are non-supersymmetric due to the presence of 
the non-trivial intersections. Thus, for example, associated to each 
of the intersections there are  massive scalar fields
which in some sense may be considered  "SUSY-partners", squarks and sleptons,
of the massless chiral fermions, have the same multiplicity $|I_{ij}|$
 and carry the same gauge quantum numbers.
The lightest of those states have  masses
\cite{afiru}
{\small \beqa
\begin{array}{cc}
t_1:  & \alpha' {\rm (Mass)}^2 =
\frac 12(-\vartheta_1+\vartheta_2+\vartheta_3) \\
t_2:  & \alpha' {\rm (Mass)}^2 =
\frac 12(\vartheta_1-\vartheta_2+\vartheta_3) \\
t_3:   & \alpha' {\rm (Mass)}^2 =
\frac 12(\vartheta_1+\vartheta_2-\vartheta_3) 
\label{tachdsix}
\end{array}
\eeqa}
Here $\vartheta_i $ are the intersection angles (in units of $\pi $)
at each of the three
subtorus. As is obvious
from these formulae  the masses depend on the angles at each intersection
and hence on the relative size of the radii.
Although  in  principle some of the scalars could  be tachyonic,
  in general it is possible
   to vary the compact radii in order  to get rid of all tachyons
of a given model (see \cite{afiru,imr}). 
Although the spectrum at the intersections has no supersymmetry, 
in the bulk of the D6-branes there is at the tree level $N=4$
SUSY gauge multiplets of each of the groups of the SM. However,
loop effects involving the fields at the intersections render the
charged SUSY partners of the gauge bosons massive and of order the
string scale \cite{imr} . Other Kaluza-Klein and winding copies of the gauge 
multiplets appear at the massive level.

Up to now we have ignored the existence or not of the Higgs system
required for the breaking of the electroweak symmetry as well as
to give masses to quarks and leptons. Looking at the $U(1)$ charges of
quarks and leptons in Table 1, we see that possible Higgs fields
coupling to quarks come in four varieties
with charges under $Q_b,Q_c$ and hypercharge given in  Table 3. 
\begin{table}[htb] \footnotesize
\renewcommand{\arraystretch}{1.25}
\begin{center}
\begin{tabular}{|c|c|c|c|}
\hline
 Higgs   &  $Q_b$  &  $Q_c$   & Y \\
\hline\hline $h_1$ & 1  &  -1 & 1/2  \\
\hline $h_2$ &   -1    &  1  &  -1/2   \\
\hline\hline $H_1$ & -1  &  -1 & 1/2  \\
\hline $H_2$ &   1    &  1  &  -1/2   \\
\hline \end{tabular}
\end{center} \caption{ Electroweak Higgs fields
\label{higgsses} }
\end{table}
Now, the question is whether for some configuration of the branes
such  Higgs fields  appear in the light spectrum.
Indeed that is the case. The $U(2)$ branes ($b,b^*$) are parallel
to the ($c,c^*$) branes along  the second torus and hence they do not intersect.
However there are open strings which stretch in between  both sets of branes
and which lead to  light scalars when the distance $Z_2$ in the second torus
is small.
 The $H_i$'s come from
the $b-c^*$ intersections whereas the $h_i$ come from the
$b-c$ intersections. 
The quadratic scalar potential for these fields has the form:
\beqa
V_2\ =\ m_H^2 (|H_1|^2+|H_2|^2)\ +\ m_h^2 (|h_1|^2+|h_2|^2)\ + \nonumber \\ 
+\ m_B^2 H_1H_2+h.c. \ +\ m_b^2h_1h_2+h.c.
\label{Higgspot}
\eeqa
The mass parameters of the potential have an
interesting  geometrical  interpretation in terms of the brane distances
and intersection angles and massless scalars appear for appropriate
values of the parameters. The number of light Higgs multiplets 
will be given by the intersection number of the (bc) and (bc*) 
branes in the first and third tori, i.e., the 
number of $h_i$ fields is given by $I_{bc}=(n_b^1m_c^1-m_b^1n_c^1)
(n_b^3m_c^3-m_b^3n_c^3)$ and the number of $H_i$ fields 
is given by $I_{bc*}=(n_b^1m_c^1+m_b^1n_c^1)
(n_b^3m_c^3+m_b^3n_c^3)$. In the example above one has $|I_{bc}|=|I_{bc*}|=1$
and there is just one Higgs sector of each type. Models in which 
only one copy of the $H_i$ or $h_i$ fields appear may be obtained.

There is no room here to describe other phenomenological aspects 
of these theories and we refer the reader to ref.\cite{imr}. Let us just
make a few general remarks. In this class of theories each quark and
lepton
resides on a different point in the six extra dimensions. Since 
in order to get Yukawa couplings the worldsheet has to stretch between
different locations for the different generations, one expects 
the natural appearance of hierarchical 
Yukawa couplings \cite{afiru2,imr} . Neutrino masses are special in this
respect. In the present models $L$ is an exact symmetry in
perturbation theory and hence only Dirac masses are allowed.
One possibility to obtain sufficiently small neutrino masses is to 
make use of the above mentioned mechanism to get hierarchies of
Yukawa couplings. Neutrino Yukawa couplings could be exponentially 
suppressed in these theories. Perhaps a more interesting possibility is the
case in which there are no Yukawa couplings at all for neutrinos. That 
is  precisely the  case if only Higgs fields of type $H_i$ above are present. 
In that case there are no perturbative masses for the lightest quarks 
nor the neutrinos. However
 there are in general dimension 6 operators of the
form  $\alpha '(L N_R)( Q_L U_R)^*$. These come from the exchange of
massive string states and are consistent with all
gauge symmetries. Plugging the u-quark  chiral condensate,
  neutrino masses of order \cite{imr} $
m_{\nu } \ \propto  (<u_Ru_L>)/M_s^2$
are obtained. For $<u_Ru_L>\propto $ $(200\ MeV)^3$ and $M_s\propto 1-10 TeV$
one gets neutrino masses of order $0.01-1 eV$'s, consistent with oscillation
experiments. 
Notice that the dimension 6 operators may have different coefficients
for different neutrino generations so there will be in general non-trivial
generation structure. We believe that the presence of Dirac neutrino masses of this
order of magnitude from this mechanism looks like a general property of low string
scale models.
 
These models are non-supersymmetric and to avoid the hierarchy 
problem one should chose the string scale to be not 
much above  the weak scale.
 As noted in ref.\cite{bgkl} , the usual
procedure \cite{aadd} 
for lowering the string scale down to 1-10 TeV while
maintaining the four-dimensional Plank mass at its experimental
value cannot be applied directly to these D6-brane toroidal models.
This is because if some of the compact radii $R_{1,2}^i, i=1,2,3$
are made large some  charged  KK modes living on the branes would become
very light.  In ref.\cite{afiru} it was
proposed a way in which one can have a low string scale compatible
with the four-dimensional large Planck mass. The idea is that the
6-torus could be small while being connected to some very large
volume manifold. For example, one can consider a region of the
6-torus away from the D6-branes, cut a ball and gluing a throat
connecting it to a large volume manifold.  In this way one would 
obtain a low string scale model without affecting directly
the brane structure discussed above.
Alternatively one can construct analogous models 
\cite{cim} in terms of 
D5-branes wrapping on $T^4\times T^2/Z_N$. In this case  the standard
mechanism to lower the string scale can be obtained by making large 
the two dimensions transverse to the D5-branes \cite{afiru,cim} .

In summary, present day knowledge of string theory in terms 
of branes allow us to follow 
a {\it bottom-up approach} to the string embedding of the SM. 
Simple  configurations of intersecting branes allow us for the first time
the construction of brane configurations with massless fermion sector
identical to that of the non-SUSY SM (see ref.\cite{csu} for a recent
attempt 
for the SUSY case). Interestingly enough, such constructions
{\it imply  the gauging of the perturbative symmetries of the SM}
like baryon and lepton number. In our opinion this provides a neat
and beautiful explanation of the surprising stability of the
proton. Furthermore, anomaly cancellation seems to work in this 
context only due to the fact that the number of colours in the SM is equal 
to the number of generations, offering a nice explanation for the generation
puzzle. 

The different sizes of the SM gauge couplings in these 
theories have to do with the different volumes on which each different D-brane
is  wrapping, and one can vary those volumes so that the observed
gauge couplings are reproduced. Thus the logarithmic unification
of couplings will be lost, like in any model with a low
string scale.  In any event, we physicists cannot rely on a single
piece of data as MSSM gauge coupling unification is.       
It could well be that the
nice agreement of gauge coupling unification in the MSSM could
be fortuitous. Recall: the apparent size of the sun 
agrees with that of the moon with a good precission.
 For centuries mankind has
given special meaning to this "size unification" 
 which turns out to be 
just an  accident in  the formation of the solar system.
 Perhaps we should learn the lesson.

\bigskip

\bigskip

\bigskip

\centerline{\bf Acknowledgements}
I am grateful to my collaborators G. Aldazabal, S. Franco, F. Marchesano,
R. Rabad\'an and A . Uranga for a very enjoyable collaboration on topics 
discussed in this talk. 
This work is partially supported by CICYT (Spain) and the   
European Commission (RTN contract HPRN-CT-2000-00148).

\end{document}